\documentclass[prl,twocolumn,showpacs,amssymb]{revtex4}

\usepackage{psfrag,graphicx}
\usepackage{dcolumn}
\usepackage{bm}
\usepackage{amsfonts,amssymb,amsmath}        
\usepackage{yfonts}
\usepackage{xcolor}

\newcommand{\be}{\begin{equation}}
\newcommand{\ee}{\end{equation}}
\newcommand{\bq}{\begin{eqnarray}}
\newcommand{\eq}{\end{eqnarray}}

\newcommand{\pp}{\boldsymbol{p}}
\newcommand{\da}{\dagger}
\newcommand{\bs}{\boldsymbol}
\begin{document}
\title{Geometric Model of Topological Insulators from the Maxwell Algebra} 

\author{Giandomenico Palumbo}
 \affiliation{Institute for Theoretical Physics, Centre for Extreme Matter and Emergent Phenomena,
Utrecht University, Princetonplein 5, 3584 CC Utrecht, The Netherlands}

 \affiliation{
Institute for Theoretical Physics, University of Amsterdam, Science Park 904, 1098 XH Amsterdam, The Netherlands}

\date{\today}

\pacs{...}

\begin{abstract}

\noindent
We propose a novel geometric model of three-dimensional topological insulators in presence of an external electromagnetic field. The gapped boundary of these systems supports relativistic quantum Hall states and is described by a Chern-Simons theory with a gauge connection that takes values in the Maxwell algebra. This represents a non-central extension of the Poincar\'e algebra and takes into account both the Lorentz and magnetic-translation symmetries of the surface states. In this way, we derive a relativistic version of the Wen-Zee term, and we show that the non-minimal coupling between the background geometry and the electromagnetic field in the model is in agreement with the main properties of the relativistic quantum Hall states in the flat space.

\end{abstract}

\maketitle

\section{I. Introduction}
It is well know that there exists a deep relation between topological phases of matter and gauge theories. In fact, at ground state, topological matter can be described by suitable topological quantum field theories \cite{Fradkin,Wen,Qi, Palumbo1}, which can be classified in terms of their topological invariants and underlying gauge groups. For instance, the interacting edge states of topological phases can be derived from suitable gauge theories \cite{Palumbo2}. Another further example is the Berry phase, which plays a crucial role in several topological systems and represents a geometric phase related to a gauge connection (Berry connection) in the momentum space \cite{Niu}. The Abelian Berry phase is given in terms of a U(1) principle bundle, formally the same one that appears in electromagnetism. This picture can be naturally extended to non-Abelian phases.\\
In recent years, there have been several efforts and proposals concerning the geometric properties of topological phases due to their importance in quantum matter. For instance, an effective geometric language has been employed to describe quantum thermal properties, such as the thermal Hall effect \cite{Read, Cappelli, Zhang, Ryu2, Stone, Palumbo3}. Novel topological responses of quantum topological fluids, such as Hall viscosity \cite{Avron,Read2}, appear when these systems are coupled to a curved background \cite{Fradkin2,Son,Kimura,Wiegmann,Gromov,You,Cappelli2}. Geometry has been also employed in the study of scattering processes in topological insulators \cite{Parente,Fukui} and in the analysis of ripples and novel emergent phenomena in graphene \cite{Vozmediano2,Gibbons,Iorio1}. Furthermore, advanced geometric theories, based on non-commutative geometry \cite{Bellissard, Susskind, Poly} and holography \cite{Zaanen,Hartnoll,McGreevy,Sachdev,Amoretti} have been employed in the study of fractional quantum Hall states, non-Fermi liquids, strongly-correlated systems, etc. (for other recent applications of curved-space formalism to condensed matter systems, see, e.g. Refs.~\cite{Vitelli,Ortix}).
Importantly, the geometric properties of quantum systems can be based on the gauge principle: the coupling between matter and spacetime is identified by gauging the global spacetime symmetries of the matter field \cite{Nakahara}. In the case of Dirac materials \cite{Bernevig,Ryu}, the quasiparticles are given by Dirac fermions and the Lorentz symmetries (i.e. rotations and boosts) emerge at ground state. Thus, in order to study these phases in curved spacetime, one has to gauge the Lorentz group and replaces the derivative with the covariant derivative, where the corresponding spin connection takes values in the Lorentz algebra \cite{Nakahara}.
The presence of the external magnetic field induces in the quantum states further symmetries, called magnetic translations, which are the proper symmetries of a (infinite) plane in presence of a constant magnetic field \cite{GMP, Cappelli3,Duval}. Although these symmetries play a crucial role in the understand of the quantum Hall fluids, such as their incompressibility, so far, they have not yet properly incorporated in any geometric model of relativistic topological phases (for the non-relativistic case, see, e.g. Refs.~\cite{Cappelli2,Haldane}).

The main goal of this paper is to present a novel geometric model of time-reversal-invariant topological insulators in three dimensions \cite{Qi,Bernevig,Ryu} that takes into account both Lorentz symmetries and magnetic translations. The corresponding geometric action on the gapped boundary, where the gap is induced by an external electromagnetic field, is implemented by gauging the Maxwell algebra \cite{Bacry,Schrader,Beckers,Lukierski}. This represents a non-central extension of the Poincar\'e algebra. It allows us to define a new effective topological field theory for the gapped boundary, given by a Chern-Simons theory with a gauge connection that takes value in the 2+1-dimensional Maxwell algebra \cite{Cangemi,Szabo,Hosein}.
The final action, written in terms of dreibein, spin connection \cite{Nakahara} and electromagnetic gauge potential contains three main elements. The first one is the standard Abelian Chern-Simons term that describes the quantum Hall conductance \cite{Wen}. The second one is given by a purely geometric contribution that describes the torsional Hall viscosity \cite{Fradkin2} and is compatible with a geometric theory recently proposed for topological superconductors \cite{Palumbo3}. Importantly, these terms define an effective exotic AdS gravitational model \cite{Witten} dual to a unitary CFT with chiral central charge $c=1$. Finally, the third one contains a novel non-minimal coupling between the Abelian gauge field and the curved background and resembles to a relativistic version of the Wen-Zee theory \cite{Wen-Zee} proposed in the study of quantum Hall fluids on a curved background. We will show that in the flat limit, our model is in agreement with the main properties of relativistic quantum Hall states.\\
The paper is structured as follows: In Sec.~II, we summarize the main properties of three-dimensional topological insulators, by focusing in their effective description in terms of Dirac Hamiltonian. Then, we derive in Sec.~III an effective geometric action by integrating out the Dirac field. By applying the holographic correspondence, we show that the gravitational theory is dual to a CFT with central charge $c=1$, which describes one-dimensional Dirac modes propagating along defect lines created on the gapped boundary.
 In Sec.~IV, we introduce the Maxwell algebra and we show that it correctly takes into account the magnetic translations, induced on the boundary of the system by the presence of an external electromagnetic field.
We show in Sec.~V that the the model found in Sec.~III can be nicely extending by gauging the Maxwell algebra. This allows us to derive new geometric terms and one of them resembles to the Wen-Zee term in quantum Hall states. We present our conclusions in Sec.~VI.
\begin{figure}[!ht]
	\centering
		\includegraphics[width=0.20\textwidth]{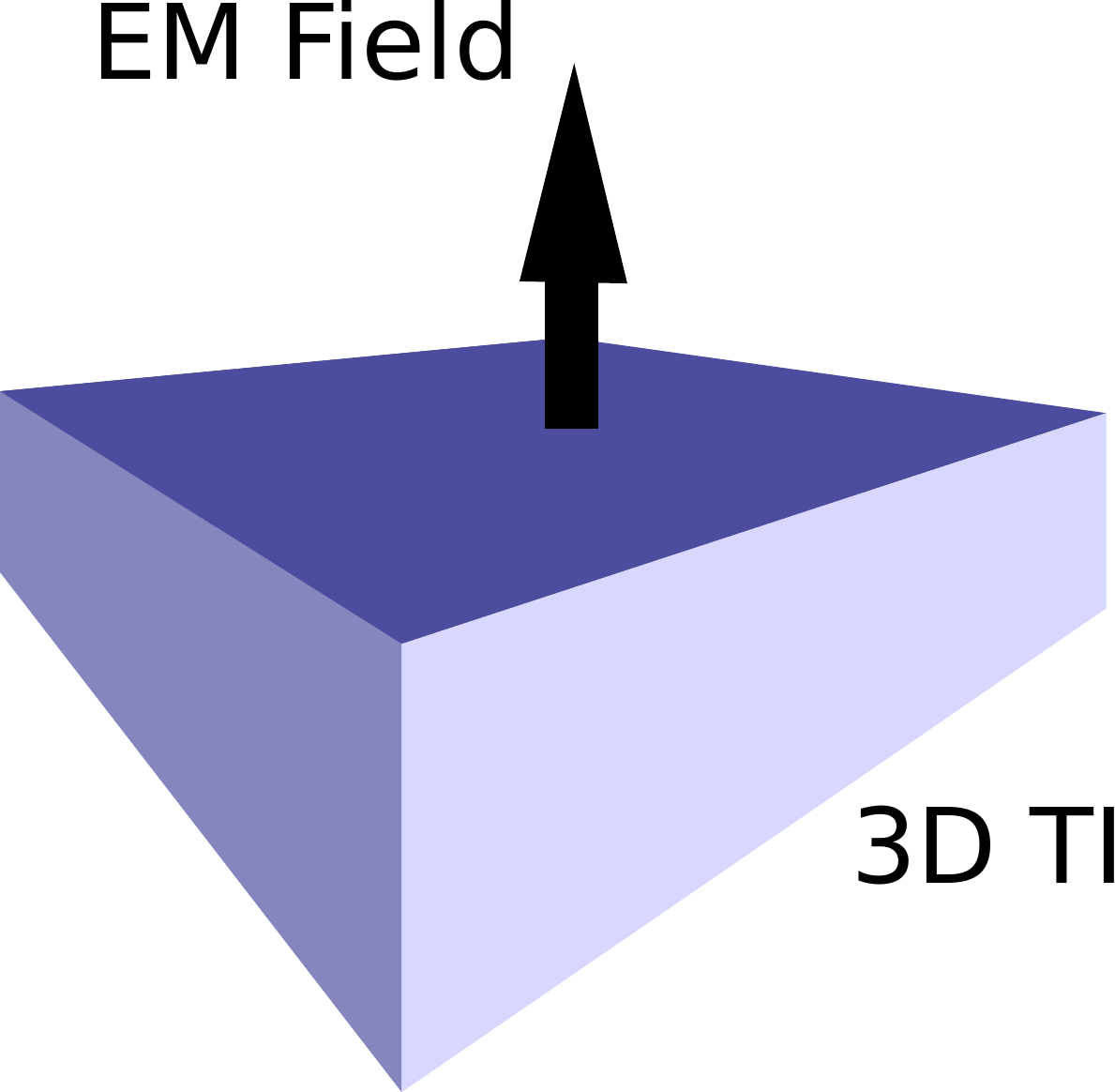}
	\caption{Schematic representation of a three-dimensional topological insulator (TI) with an external electromagnetic (EM) field represented by an arrow. This gauge field opens a gap on the boundary of the TI, i.e. the two-dimensional Dirac modes acquire a mass and the translation symmetries are replaced by magnetic-translation symmetries, which characterize the properties of the relativistic quantum Hall states induced on the boundary.}
	\label{Fig1}
\end{figure}
\section{II. Three-dimensional topological insulators}
We start by summarizing the main properties of three-dimensional topological insulators in the quantum-field-theory framework. The microscopic Hamiltonian is $H=\sum_{\pp}\bs{\psi}_{\pp}^{\da}h(\pp)\bs{\psi}_{\pp}$, where $\pp\in[0,2\pi)\times[0,2\pi)\times[0,2\pi)$ and $h(\pp)$ is a generic kernel Hamiltonian belonging to the class AII of free-fermion models \cite{Ryu,Qi}.
It is well known that the continuum real-space Hamiltonian is a Dirac Hamiltonian
\begin{eqnarray} \label{Hamiltonian}
H =\int d^3x \,\, \psi^{\dagger}(i\xi^{j}\partial_{j}+i\,\zeta\, m)\psi,
\end{eqnarray}
where $j=1,2,3$, $\xi^{j}=\sigma^{1} \otimes \sigma^{j}$, $\zeta =\sigma^{3} \otimes \mathbb{I}_{2 \times 2}$, $\mathbb{I}_{2 \times 2}$ is the identity matrix, $\sigma^{j}$ are the Pauli matrices, $m$ is the Dirac mass and $\psi$ is a four-component spinor. Due to the charge conservation in topological insulators, we can study their topological response to the presence of an external electromagnetic field $A_{\mu}$, see Fig.1. Thus, the corresponding fermionic action defined on a Lorentzian manifold $M$ can be written as follows
\begin{eqnarray}
\label{Diracflat}
S^\text{3D}[\psi,\overline{\psi}]=\int_{M} d^{4}x\,\,
\overline{\psi}\,(i\,\gamma^{\mu}\partial_{\mu}+\gamma^{\mu}A_{\mu}-m)\psi,
\end{eqnarray}
where $\mu=0,1,2,3$, $\overline{\psi}=\psi^{\dagger}\gamma^{0}$, and $\gamma^{j}=\gamma^{0}\xi^{j}$ are the Dirac matrices. Because we are considering a U(1) dynamical gauge field, we have to add up to the above action the Maxwell term
\begin{eqnarray}
\label{Maxwell}
S^\text{M}[A_{\mu}]=\frac{1}{4}\int_{M} d^{4}x\,F^{\mu\nu}F_{\mu\nu},
\end{eqnarray}
where $F_{\mu\nu}=\partial_{\mu}A_{\nu}-\partial_{\nu}A_{\mu}$ is the Faraday tensor.
In order to derive the topological effective theory that describes the three-dimensional topological insulator in the low-energy limit, 
we integrate out the spinor field in the partition function of $S^\text{3D}[\psi,\overline{\psi}, A_{\mu}]$. The corresponding effective action $S^\text{3D}_{\rm eff}[A_{\mu}]$ defined by 
\begin{eqnarray}
e^{i S^\text{3D}_{\rm eff}[A_{\mu}]}=\int D\overline{\psi}\,\textit{D}\psi\,e^{i S^\text{3D}[\psi,\overline{\psi}, A_{\mu}]},
\end{eqnarray}
has always a topological term that becomes dominant at low energy and describes the properties of the ground state. This term in 3+1 dimensions is proportional to the axion topological term and the final effective bosonic action is simply given by
\begin{eqnarray}
\label{EFT}
S=\frac{1}{4}\int_{M} d^{4}x\,\left(F^{\mu\nu}F_{\mu\nu}+\frac{\theta}{8\pi^{2}}\epsilon^{\mu\nu\alpha\beta}\,F_{\mu\nu}F_{\alpha\beta}\right),
\end{eqnarray}
where $\theta=\pi$ in the case of topological insulators in the non-trivial $Z_{2}$ topological phase \cite{Qi}. Importantly, only when $\theta=0, \pi$ the time-reversal symmetry in the bulk is respected by the effective action (\ref{EFT}). However, the presence of an electromagnetic field breaks the time-reversal symmetry on the boundary and generate a mass term in the two-dimensional helical Dirac modes. These massive modes coupled to $A_{\mu}$ are represented by a 2+1-dimensional massive Dirac action similarly to that one expressed in Eq.~(\ref{Diracflat}). In this case the corresponding topological effective action is given by Chern-Simons theory that can be simply derived by employing the Stokes theorem on the topological axion term in Eq.~(\ref{EFT}). In fact, this is a total derivative that gives rise to an Abelian Chern-Simons term on the boundary. This topological quantum field theory describes the Hall conductance of relativistic quantum Hall states on the surface \cite{Qi}.
For simplicity, we consider the system defined on a Lorentzian manifold $M=R\times \Sigma$ where space-like part $\Sigma$ has periodic boundary conditions in $x$ and $y$, while the periodicity in $z$ is broken in order to have a boundary made by two disconnected surface.
Similarly to the case of topological superconductors \cite{Palumbo4}, here the whole gapped boundary can be seen an effective two-dimensional topological phase belonging the class A \cite{Ryu} and supports topological protected effective edge modes. These one-dimensional Dirac edge modes are trapped by the defect lines that can be created on the gapped surfaces by employing a couple of local Zeeman fields \cite{Palumbo5}. In the next section, we will show that the holographic correspondence, already employed in the case of topological superconductors \cite{Palumbo3}, is still valid in the case of topological insulators and allows us to derive the right value of the chiral central charge associated to the chiral Dirac modes.

\section{III. Holographic correspondence in topological insulators}
In order to derive an effective geometric theory for the three-dimensional topological insulator, we first observe that the Dirac theory is invariant under the global Poincar\'e group. In other words, in the low-energy limit, topological insulators support further emergent relativistic symmetries, given by Lorentz boosts and rotations, and spacetime translations. By gauging these symmetries, we can understand how the system behaves under local geometric transformations that preserve locally the Poincar\'e group. This approach has been also employed in the study of geometric defects in topological phases and in the generalization of the Luttinger theory \cite{Luttinger}, where a minimal coupling between fermions and the background geometry has been used in order to derive thermal quantum effects, such as the thermal Hall effect \cite{Read, Cappelli, Zhang, Ryu2, Stone, Palumbo3}. The gauging procedure in the fermion theory is made by replacing the standard derivative with a covariant derivative and by introducing the tetrads, which allows us to write the Dirac action in the curved space $M_{c}$,
given by 
\begin{eqnarray}\label{Dirac2}
S^\text{3D}[\psi,\overline{\psi}, A_{\mu}, \omega_{\mu}, e_{a}^{\mu}]=\int_{M_{c}} d^{4}x\,|e|\,
\overline{\psi}\,(i\,\widehat{\gamma}^{\mu}D_{\mu}-m)\psi,
\end{eqnarray}
where $|e|$ is the determinant of the tetrads $e_{a}^{\mu}$, $\widehat{\gamma}^{\mu}=e_{a}^{\mu}\gamma^{a}$ and $D_{\mu}=i\partial_{\mu}+A_{\mu}+\omega_{\mu}$, where $\omega_{\mu}$ is the spin connection \cite{Nakahara}. Clearly, in the flat limit, we recover the (\ref{Diracflat}), because in that case $|e|=1$, $\omega_{\mu}=0$ and $e_{a}^{\mu}\gamma^{a}=\delta_{a}^{\mu}\gamma^{a}=\gamma^{\mu}$. Importantly, we are interested to a torsion-full spin connection that is able to take into account also possible dislocations in the system. \cite{Fradkin2,Vozmediano,Zaanen2}. More important, this choice is compatible with the fact that in the gauge-theory language, $\omega_{\mu}$ and $e_{a}^{\mu}$ are independent fields, with the former related to the Lorentz symmetries and the latter related to the spacetime translations.
This has important implications in the derivation of the topological effective theory in curved space. By integrating out fermions, we have that \cite{Zanelli}
\begin{align}\label{Zanelli}
& S_\text{top}^\text{3D}[A_{\mu},\omega_{\mu},e_{\mu}]= \frac{1}{32\pi}\int_{M_{c}} d^{4}x\,\epsilon^{\mu\nu\alpha\beta}\,F_{\mu\nu}F_{\alpha\beta}\,-\nonumber \\
& k\,\int_{M_{c}} d^{4}x\,
\epsilon^{\mu\nu\alpha\beta}\, \text{tr}\, \Big[ R_{\mu\nu}R_{\alpha\beta}+ 
{1\over \eta^{2}}\Big(T_{\mu\nu}T_{\alpha\beta}-R_{\mu\nu}e_{\alpha}e_{\beta}\Big)\Big],
\end{align}
where $k=\frac{1}{192\pi}$ and $\eta$ is a dimensionful parameter related to the Hall viscosity \cite{Fradkin2, Hughes}. Here, $R_{\mu\nu}=R_{\mu\nu}^{ab}\,i\,[\gamma_{a},\gamma_{b}]/4$ and $T_{\mu\nu}=\frac{i}{2}\gamma_{a}T^{a}_{\mu\nu}$, are the Riemann and torsion tensor, respectively given by 
\begin{align}
R_{\mu\nu}^{ab} & =\partial_{\mu} \omega_{\nu}^{ab}-\partial_{\nu}\omega_{\mu}^{ab}+\omega_{\mu\,\,c}^{a}\omega_{\nu}^{cb}-\omega_{\nu\,\,c}^{a}\omega_{\mu}^{cb}, \nonumber \\
T^{a}_{\mu\nu} & =\partial_{\mu}e_{\nu}^{a}-\partial_{\nu}e_{\mu}^{a}+\omega_{\mu\,\, b}^{a}e_{\nu}^{b}-\omega_{\nu\,\, b}^{a}e_{\mu}^{b}. \hspace{0.5cm}
\end{align}
In the action (\ref{Zanelli}), we recognize the Pontryagin invariant \cite{Eguchi} in the second term, while the third term is proportional to the Nieh-Yan invariant \cite{Nieh}. However, this is not the conclusive topological action because, so far, we have omitted possible contributions in the effective action induced by the Maxwell term. In fact, the Maxwell theory in 3+1 dimensions supports the duality symmetry, defined by the following transformation
\begin{align}
F_{\mu\nu}\rightarrow F'_{\mu\nu}=F_{\mu\nu}\, \cos\, \theta+\widehat{F}_{\mu\nu}\, \sin\, \theta
\end{align}
where $\theta$ is an angle and $\widehat{F}_{\mu\nu}=\frac{1}{2}\epsilon_{\mu\nu\alpha\beta}F^{\alpha\beta}$. However, this symmetry holds only in the flat case and is broken by quantum effects when the theory is defined on a generic curved spacetime, as shown in Refs.~ \cite{Reuter,Dolgov,Agullo}. It is possible to show that the duality anomaly induces a topological term proportional to the Pontryagin invariant, similarly to that one found in the fermionic sector. Thus, We replace $k$ with $k'=2k$ in Eq.~(\ref{Zanelli}), in order to define the total effective topological theory. Notice, that the contribution of the torsion term proportional to $1/\eta^{2}$ is not really relevant in the holographic correspondence, as we will see below.

We can now easily derive the two-dimensional effective theory $S_{\rm top,k}^\text{2D}$ on the gapped boundary by employing the Stokes theorem due to the fact that $S_{\rm top}^\text{3D}$ is a total derivative. We find that
\begin{align}
\label{DoubleCS}
& S_{\rm top,k}^\text{2D}[\omega_{\mu},e_{\mu},A_{\mu}] = \frac{2}{8\pi}\int d^{3}x\,\epsilon^{\mu\nu\lambda}A_{\mu}\partial_{\nu}A_{\lambda}\,-\nonumber \\  
& 2 k'\,\int d^{3}x\, \epsilon^{\mu\nu\lambda}
 \text{tr} \left(\omega_{\mu}\partial_{\nu}\omega_{\lambda}+\frac{2}{3}\,\omega_{\mu}\omega_{\nu}\omega_{\lambda}+\right.  \left.\frac{1}{\eta^{2}}\,T_{\mu\nu}e_{\lambda}\right),
\end{align}
where the factor $2$ in front of both terms takes into account both the two disconnected surfaces.
The second line in this action describes the exotic AdS gravity \cite{Witten}, which, differently from the standard Einstein-Hilbert theory, breaks parity and time-reversal symmetry. This is agreement with the symmetries of the gapped boundary and allows us to employ the holographic correspondence in order to derive the corresponding topological chiral central charge \cite{Palumbo3}. It is possible to derive its value by analyzing the holographic stress-energy tensor, defined on the (asymptotic) boundary of AdS$_{2+1}$ \cite{Kraus}.
It can be shown that the holographic stress-energy tensor $\tau^{uv}=\tau^{u}_{i}e^{i v}$ is not symmetric,
\begin{eqnarray}\label{Lorentz1}
\tau^{[uv]} =\tau^{uv}-\tau^{vu}= \frac{2\pi\,k'}{|e|}\epsilon^{ij}\,R_{ij}^{uv},
\end{eqnarray}
where $R_{ij}^{uv}$ ($u,v=0,1$) is the Riemann tensor defined on the asymptotic boundary.
The failure of $\tau^{[uv]}$ to be symmetric implies that 
on the (1+1)-D asymptotic boundary there is a gravitational (Lorentz) anomaly \cite{Cappelli}. This quantum anomaly appears in the dual chiral CFT and is proportional to the chiral central charge $c$ \cite{Cappelli, Stone}
\begin{eqnarray}\label{Lorentz}
\tau_\text{CFT}^{[uv]} = \frac{c}{48 |e|}\,\epsilon^{ij}\,R_{ij}^{uv}.
\end{eqnarray}
Due to the AdS$_{2+1}$/CFT$_{1+1}$ correspondence, Eqs.~(\ref{Lorentz1}, \ref{Lorentz}) imply that \cite{Blagojevic, Klemm}
\begin{eqnarray}
c=192\pi\,k=1. 
\end{eqnarray}
In other words, the CFT describes a single one-dimensional Dirac mode trapped by defect lines created on the gapped boundary of the three-dimensional system. We can also see these defects as an effective boundary of the two-dimensional relativistic quantum Hall fluid.

\section{IV. Magnetic translations and the Maxwell algebra}
In the previous section, we have seen that a geometric model is compatible with the main properties of a three-dimensional topological insulator with gapped boundary. However, in presence of an external magnetic field, there are further symmetries, called magnetic translations $t_{u}$ \cite{GMP, Cappelli3,Bernevig2}, that have not yet encoded in the geometric model.
It is well known that in the (relativistic) quantum Hall states, the ordinary spacial translations are replaced by $t_{u}$, which are the proper symmetries of an (infinite) plane in presence of a constant magnetic field.
They are defined as follows
\begin{align}
t_{u}=e^{i\,u_{a}K^{a}},
\end{align}
where $u=\{u_{a}\}=\{u_{x},u_{y}\}$ is the finite translation vector, such that $t_{u}:\, x_{a}\rightarrow x_{a}'=x_{a}+u_{a}$, while the generators $K_{a}$ are given by
\begin{align}\label{MT}
K_{a}=p_{a}-A_{a},  \hspace{0.5cm} [K_{a}, K_{b}]=\frac{i}{l^{2}}\,\epsilon_{ab},
\end{align}
where $p_{a}$ are the standard momenta that commute, i.e. $[p_{a},p_{b}]=0$ (to not confuse with the operators in Eq.~ (\ref{commutators})), $A_{a}$ is the gauge potential and $l=1/\sqrt{q\,B}$ is the magnetic length, with $B$ the magnetic field and $q$ the electric charge. At this point, it can be easily shown that
\begin{align}\label{GMP}
[t_{u},t_{v}]=-2 i\,\sin \left(\frac{1}{2\,l^{2}}\, u \wedge v\right) t_{u+v}.
\end{align}
This defines the magnetic translation algebra, also known as Girvin-Plazmann-MacDonald (GMP) algebra \cite{GMP}, which is at the base of the area-preserving diffeomorphisms on the plane \cite{Cappelli3,Kogan}, which explains the incompressibility of quantum Hall fluids. Moreover, the area-preserving-diffeomorphisms is mathematically described through the $W_{\infty}$ algebra, which allows to derive also the Wen-Zee term, as shown in Ref.~\cite{Cappelli2}. Thus, at least in non-relativistic topological phases, the Wen-Zee theory is related to the area-preserving diffeomorphisms and magnetic translations. What is the situation in the case of relativistic systems? \cite{Son2}

In the next section, we will propose a novel geometric model by considering the Maxwell algebra as fundamental algebra of the relativistic quantum Hall states. This algebra represents a non-central extension of the 2+1-dimensional Poincar\'e algebra and takes into account the presence of an electromagnetic field in the Minkowski spacetime \cite{Cangemi,Szabo,Hosein}. This allows us to fully encode the magnetic translations in our geometric model. At a more formal level, the adoption of the Maxwell algebra represents a modification of the tangent bundle in contrast with the more conventional approach (employed also in the previous section) where one introduces a U(1) principle bundle \cite{Nakahara}.  The Maxwell algebra in 2+1 dimensions is then defined by the following commutators
\begin{align}\label{commutators}
[P_{a},P_{b}]=\epsilon_{abc}Z^{c},\hspace{0.2cm} [J_{a},J_{b}]=\epsilon_{abc}J^{c}, \hspace{0.2cm} [J_{a},P_{b}]=\epsilon_{abc}P^{c}, \nonumber \\ [J_{a},Z_{b}]=\epsilon_{abc}Z^{c},\hspace{0.2cm} [P_{a},Z_{b}]=[Z_{a},Z_{b}]=0, \hspace{0.8cm}
\end{align}
where $a,b,c=0,1,2$, $P_{a}$ are the generators of spacetime translations, $J_{a}=\frac{1}{2}\epsilon_{abc}J^{bc}$ are the dual generators of Lorentz rotations and boosts and $Z_{a}$ are the new generators of the Maxwell algebra. The corresponding (internal) invariant metric $\Omega_{AB}=\langle X_{A}, X_{B}\rangle$, with $X_{A}=(P_{a},J_{a}, Z_{a})$ and $A,B=0,..,8$, is identified through the following relations
\begin{align}
\langle P_{a},P_{b}\rangle=\gamma\,\eta_{ab},\hspace{0.2cm} \langle J_{a},J_{b}\rangle=\alpha\, \eta_{ab}, \hspace{0.2cm} \langle J_{a},P_{b}\rangle=0, \nonumber \\ \langle J_{a},Z_{b}\rangle=\gamma\,\eta_{ab},\hspace{0.2cm} \langle P_{a},Z_{b}\rangle=\langle Z_{a},Z_{b}\rangle=0, \hspace{0.8cm}
\end{align}
where $\alpha$ and $\gamma$ are real parameters. Importantly, this algebra is not semi-simple and this implies that the matrix form of $\Omega^{AB}$ (with $\Omega_{AC}\Omega^{BC}=\delta_{A}^{B}$) does not coincide with that one of $\Omega_{AB}$ \cite{Witten2}. In other words, the Casimir operator $W=\Omega_{AB}X^{A}X^{B}$ is not equivalent to $W'=\Omega^{AB}X_{A}X_{B}$, which has an important implication in the construction of the gauge invariant action as we will see in the next section.

We can easily see that the Maxwell algebra takes into account both the Lorentz symmetries and the (spatial) magnetic translations of the relativistic quantum Hall states. For simplicity, we avoid the Lorentz rotations and boosts and we focus on the first spatial anti-commuting relations defined in Eq.~ (\ref{commutators}), i.e.
\begin{align}
[P_{a},P_{b}]=\epsilon_{ab} Z^{0}.
\end{align}
With the following identifications
\begin{align}
P_{a}\equiv l\,K_{a}, \hspace{0.3cm}
Z^{0}\equiv i,
\end{align}
we exactly recover the anti-commuting relations in Eq.~(\ref{MT}) and consequently we find the magnetic translation algebra defined in Eq.~ (\ref{GMP}).

\section{V. Generalized Wen-Zee term and Hall viscosity}
As we have seen in Sec. III, the U(1) gauge field and the curved background do not mix at the level of the effective action. This is in contract with the situation in non-relativistic systems, where the Wen-Zee term characterizes the topological states defined on curved surfaces \cite{Wen-Zee}. In that case, the Abelian spin connection $\widehat{\omega}_{\lambda}$ is coupled to the electromagnetic gauge field through the Wen-Zee term
\begin{align}\label{WZ}
S_{\rm WZ} = 
\frac{\nu \bar{s}}{4\pi}\int d^{3}x\, \epsilon^{\mu\nu\lambda}
 A_{\mu}\partial_{\nu}\widehat{\omega}_{\lambda},
\end{align}
where $\nu$ is the filling factor and $\bar{s}$ is the shift. This action in a torsionless background describes the Hall viscosity in term of the parameter $\bar{s}$, which corresponds to the intrinsic angular momentum of the low-energy excitations of the quantum Hall fluid. In relativistic topological systems, a relativistic version of Wen-Zee term is not obvious, mainly because in these cases the spin connection is non-Abelian and the Hall viscosity is given in terms of the torsional viscosity \cite{Fradkin2} (note that a relativistic version of this term has been proposed so far only in the hydrodynamic approach \cite{Son2}). 

We now show that a relativistic Wen-Zee term on the gapped boundary can be consistently derived by gauging the 2+1-dimensional Maxwell algebra introduced in the previous section. The gauge connection $\mathcal{A_{\mu}}$ that takes values in this algebra and is given by
\begin{eqnarray}
\mathcal{A_{\mu}}=\mathcal{A}^{A}_{\mu}X_{A}=\frac{1}{\beta}\,e_{\mu}^{a}P_{a}+\omega_{\mu}^{a}J_{a}+\widehat{A}_{\mu}^{a}Z_{a},
\end{eqnarray}
where $\beta$ is a dimensionful parameter and $\widehat{A}_{\mu}^{a}$ are three Abelian gauge fields. In terms of the curvature tensor $\mathcal{F_{\mu\nu}}$, we find
\begin{eqnarray}
\mathcal{F_{\mu\nu}}=\mathcal{F_{\mu\nu}}^{A}X_{A}=T_{\mu\nu}^{a}P_{a}+R_{\mu\nu}^{a}J_{a}+\widehat{F}_{\mu\nu}^{a}Z_{a},
\end{eqnarray}
where $T_{\mu\nu}^{a}$ and $R_{\mu\nu}^{a}$ are the torsion and Riemann tensor, respectively, while
\begin{eqnarray}
\widehat{F}_{\mu\nu}^{a}=\partial_{\mu}\widehat{A}_{\nu}^{a}-\partial_{\nu}\widehat{A}_{\mu}^{a}+\epsilon_{bc}^{a}\left(\frac{1}{\beta^{2}}\, e^{b}_{\mu}\,e^{c}_{\nu}+\omega_{\mu}^{b}\widehat{A}_{\nu}^{c}+\widehat{A}_{\mu}^{b}\omega_{\nu}^{c}\right).
\end{eqnarray}
The most general Chern-Simons action based on the Maxwell algebra is then given by
\begin{align}\label{MaxCS}
S_{\rm CS}[\mathcal{A_{\mu}}]=\hspace{3.0cm} \nonumber \\ -\frac{1}{4\pi}\int d^{3}x\, \epsilon^{\mu\nu\lambda}\left[\left(\frac{1}{2}\Omega_{AB}\mathcal{A}^{A}_{\mu}\mathcal{F}_{\nu\lambda}^{B}-\frac{1}{3}\Omega_{CD}f^{D}_{AB}\mathcal{A}^{A}_{\mu}\mathcal{A}^{B}_{\nu}\mathcal{A}^{C}_{\lambda}\right)\right.\nonumber\\
+\left.\left(\frac{1}{2}\Omega^{AB}\mathcal{A}_{\mu A}\mathcal{F}_{\nu\lambda B}-\frac{1}{3}\Omega^{CD} f_{D}^{AB}\mathcal{A}_{\mu A}\mathcal{A}_{\nu B}\mathcal{A}_{\lambda C}\right)\right],
\end{align}
where $f^{D}_{AB}$ are the structure constants, such that $[X_{A},X_{B}]=f_{AB}^{D}X_{D}$ (a similar relation holds for $f_{D}^{AB}$).
Let us now expand the above actions in terms of the physical gauge fields. In order to simplify the notation, here we employ the form formalism in the final topological action
\begin{align}\label{finalCS}
S_{\rm CS}[e,\omega, \widehat{A}]= \int \rm tr [\varrho_{1}CS(\omega)+\varrho_{2}\,e\wedge D_{\omega}e+ \nonumber \\  \varrho_{3}\widehat{A}\wedge (R+\varrho_{4}\,e\wedge e)+\varrho_{5}\widehat{A}\wedge D_{\omega}\widehat{A}],
\end{align}
with
\begin{align}
{\rm CS}(\omega)=\omega \wedge d\omega+\frac{2}{3}\,\omega\wedge \omega\wedge \omega.
\end{align}
The trace is taken on the gauge index, $D_{\omega}=d+\omega$ is the exterior covariant derivative, such that $T=D_{\omega}e$, and $\varrho_{i}$ are functions of the parameters $\alpha$, $\beta$ and $\gamma$, namely
\begin{align}
\varrho_{1}=-\frac{\alpha}{4\pi}, \hspace{0.2cm} \varrho_{2}=-\frac{1}{4 \pi\beta^{2}}\left(\gamma+\frac{1}{\gamma}\right), \hspace{0.2cm} \varrho_{3}=-\frac{1}{2\pi}\left(\gamma+\frac{1}{\gamma}\right), \nonumber \\
\varrho_{4}=\frac{\alpha}{4\pi\beta^{2} \gamma^{2}}, \hspace{0.3cm} \varrho_{5}=\frac{\alpha}{4\pi\gamma^{2}}. \hspace{2.6cm}
\end{align}
By comparing these parameters with those ones in Eq.~(\ref{DoubleCS}), we find $\alpha=\gamma^{2}=1/12$, $\beta^{2}=26\sqrt{3}\,\eta^{2}$, and $\eta^{2}$ is proportional to the inverse of the torsional Hall viscosity.

Bear in mind that so far we have dealt with three independent U(1) gauge fields $\widehat{A}_{\mu}^{a}$ even if only one can be associated to the physical electromagnetic field. This is identified with 
\begin{align}\label{gauge}
\widehat{A}_{\mu}^{0}\equiv A_{\mu}, 
\end{align}
while $\widehat{A}_{\mu}^{1}$ and $\widehat{A}_{\mu}^{2}$ can be seen as auxiliary fields. Notice that if we fix to zero these two gauge fields, then we lose the gauge invariance of the theory. At this point, several comments about the action in Eq.~(\ref{finalCS}) are necessary in order to clarify and strength the main properties of our model.\\
Firstly, the Chern-Simons action in Eq.~(\ref{finalCS}) represents a natural but non-trivial generalization of the effective theory found in Eq.~(\ref{DoubleCS}). Moreover, we have found also a relativistic version the Wen-Zee term (\ref{WZ}) given by the third term in Eq.~(\ref{DoubleCS}). \\
Secondly, the torsional Hall viscosity derived in Ref.~\cite{Fradkin2} remains unchanged here. In fact, we can calculate the stress-energy tensor $\mathcal{T}_{a}^{\mu}$ by varying the action with respect to the dreibein $e_{\mu}^{a}$
\begin{align}\label{Stress-Energy}
\mathcal{T}_{a}^{\mu}=\frac{\delta S_{\rm CS}}{\delta\,e^{a}_{\mu}}=-\epsilon^{\mu\nu\lambda}\left(\frac{1}{24 \pi\eta^{2}}T_{a\nu\lambda}+\frac{1}{48 \pi^{2}\eta^{2}}\epsilon_{abc}\,\widehat{A}^{b}_{\nu}e^{c}_{\lambda}\right).
\end{align}
Here, we find an extra contribution to the stress-energy tensor given by the above second term. However, both terms are proportional to the inverse of $\eta^{2}$, which is only dimensionful parameter and is related to the torsional Hall viscosity. Thus, our relativistic Wen-Zee term does not contribute to the Hall viscosity because $\omega$ and $e$ are independent fields due to the non-null torsion.
Importantly, the torsional Hall viscosity appears also in the flat limit as we see now. The current $\mathcal{J}^{\mu}$ associated to the variation of the action with respect to the electromagnetic gauge field $A_{\mu}$ in the Minkowski spacetime (i.e. $\omega_{\mu}=0$ and $e_{\mu}^{a}=\delta_{\mu}^{a}$), is given by
\begin{eqnarray}\label{Hall}
\mathcal{J}^{i}=\left.\frac{\delta S_{\rm CS}}{\delta\,A_{i}}\right|_{\rm flat}=\epsilon^{ik}\frac{1}{2\pi}E_{k},
\end{eqnarray}
with $i,k=x,y$ and
\begin{eqnarray}\label{GaussCS}
\mathcal{J}^{0}=\left.\frac{\delta S_{\rm CS}}{\delta\,A_{0}}\right|_{\rm flat}=\frac{1}{2\pi}(B+B_{0}),
\end{eqnarray}
where $B_{0}=-\frac{1}{24\pi\eta^{2}}$, and $E_{k}$ and $B$ represent the electric and magnetic field, respectively. Eq.~(\ref{Hall}) represents the standard quantum Hall law, while Eq.~(\ref{GaussCS}) is the Chern-Simons Gauss law \cite{Frohlich} and contains a novel contribution induced by the torsional Hall viscosity. In the flat limit, this can see as an effective constant magnetic field $B_{0}$ for a simple reason. As shown in Ref.~\cite{Ryu3}, the torsional Hall viscosity for the gapped boundary of three-dimensional topological systems, is proportional to the external constant magnetic (Zeeman) field that induces a Dirac mass in the surface states. Moreover, always in the flat background, the fourth term in the action (\ref{finalCS}) that depends on the electromagnetic field, reduces to a chemical potential term, namely
\begin{eqnarray}
\left.\frac{1}{96 \pi^{2}\eta^{2}}\int d^{3}x\, \epsilon^{\mu\nu\lambda}\epsilon_{0ab}\,e_{\mu}^{a}A_{\nu}e^{b}_{\lambda}\right|_{\rm flat}\longrightarrow\nonumber \\ \longrightarrow \int d^{3}x\, \kappa A_{0}\equiv \int d^{3}x\, I^{\mu}A_{\mu}^{\rm ext},
\end{eqnarray}
where $\kappa=\frac{1}{2\pi}B_{0}$ is the chemical potential, while
\begin{eqnarray}
 I^{\mu}=\frac{1}{2\pi}\,\epsilon^{\mu\nu\lambda}\partial_{\nu}A_{\lambda}, \hspace{0.3cm} A_{i}^{\rm ext}=\frac{B_{0}}{2}\epsilon_{i j}x^{j},
\end{eqnarray}
are the topological current and an external gauge potential, respectively. This exactly reproduces the constant magnetic background in the quantum Hall fluids and is also in agreement with the chemical potential term derived in Ref.~\cite{Tong}.
We then conclude that our geometric model is compatible with the main properties of the relativistic quantum Hall states.

We finally speculate on the possible interpretation of our (charged) geometric model in the holographic context. It is not an easy task to deal with a non-minimal coupling between geometry and gauge fields. Formally, each Abelian field is associated to a U(1) gauge group, which is isomorphic to special orthogonal group SO(2). By performing a Wigner-Inonu contraction \cite{Gilmore} on this group, we have SO(2)$\rightarrow \mathcal{E}(1)$, where $\mathcal{E}(1)$ is the Euclidean group that describes translations in one dimension. In this way, the three U(1) gauge groups can be related to three independent translations. In the gauge-theory language this implies that the gauge fields $\widehat{A}_{\mu}^{a}$ induce a dreibein built by gauging the three $\mathcal{E}(1)$ algebras. We can possibly write
\begin{eqnarray}\label{GaussCS}
\widehat{A}_{\mu}^{a}\rightarrow \Theta\, e^{a}_{\mu},
\end{eqnarray}
where $\Theta$ is a coefficient that we assume to be constant.
With the above replacement, the action in Eq.~(\ref{finalCS}) becomes purely geometric, including now also an Einstein-Hilbert term and a cosmological constant $\Lambda \propto \varrho_{4}$. It coincides with the Mielke-Baekler action \cite{Mielke}, which is a model that has been already studied in the holographic context and the corresponding Virasoro central charges have been derived \cite{Blagojevic,Klemm}. As we have shown in the Sec. III, only the chiral central charge $c$ is related to the topological phase. Thus, in this generalized geometric model, we recover $c=1$.

\section{VI. Conclusions}
In this work, we have proposed a new geometric model of topological insulators based on the Maxwell algebra. This is a non-central extension of the Poincar\'e algebra that takes into account the symmetries of the gapped boundary states, i.e. the Lorentz symmetries and magnetic translations.
The Chern-Simons theory that describes these states is built in terms of a gauge connection that takes values in the Maxwell algebra. The standard U(1) Chern-Simons theory is consistently reproduced in this model together with gravitational terms and two novel ones that represent a non-minimal coupling between the electromagnetic field and the curved background. We have shown that the purely gravitational part of the theory is compatible with the presence of one-dimensional Dirac modes propagating along the defect lines created on the gapped boundary. The corresponding CFT with chiral central charge $c=1$ has been derived through the holographic correspondence. Importantly, our approach can be applied also to topological phases, such as two-dimensional Chern insulators, where the magnetic translations (under suitable geometric conditions) occur without the presence of any external electromagnetic field \cite{Roy1,Roy2}. In conclusion, our theory opens the way to the application of Maxwell geometry in topological phases of matter.

\section{Acknowledgments}
We thank Andrea Cappelli, Jan Zaanen, Hai-Qing Zhang and Enrico Randellini for comments and suggestions. This work is part of the DITP
consortium, a program of the Netherlands Organisation
for Scientific Research (NWO) that is funded by the Dutch
Ministry of Education, Culture and Science (OCW).

\end{document}